\documentclass{article} 
\usepackage{nips13submit_e,times}
\usepackage{hyperref}
\usepackage{url}
\usepackage{amsmath}
\usepackage{amsfonts}
\usepackage[pdftex]{graphicx} 

\newcommand{\+}{\,+\,}
\renewcommand{\-}{\,-\,}
\renewcommand{\v}[1]{\boldsymbol{#1}}

\newcommand{\half}{\frac 1 2 \,}
\newcommand{\const}{\mathrm{const}}
\newcommand{\diag}[1]{\mathrm{diag} \left( #1 \right) }
\newcommand{\h}{\v h}
\newcommand{\xt}{\v x_t}
\newcommand{\Ch}{\v{C}}
\renewcommand{\th}{\v \theta}
\newcommand{\thhat}{\v {\hat \theta}}
\newcommand{\LJ}{\mathcal{LJ}}

\newcommand{\Cthhatinv}{\v C(\thhat)^{-1}}
\newcommand{\nuh}{\v{\hat \nu}}
\newcommand{\Ek}[1]{\mathbb{E}_{q(\v k)} \left[ #1 \right]}

\newcommand{\Ehth}[1]{\mathbb{E}_{q(\v h, \th)} \left[ #1 \right]}

\newcommand{\hstar}{\v {h^*}}
\newcommand{\Lstar}{\v {L^*}}
\newcommand{\Lstarinv}{ \left( \v {L^*} \right)^{-1} }
\newcommand{\jacob}[2]{\frac {\partial #1} {\partial #2} }

\newcommand{\hesscross}[3]{\frac {\partial^2 #1} {\partial #2 \, \partial #3} }
\newcommand{\LP}{\mathcal{LP}}
\newcommand{\LPhstar}{\LP_{\hstar}}

\newcommand{\hhatstar}{\v {\hat h^*}}
\newcommand{\Lamstar}{\v {\Lambda^*}}
\newcommand{\hhat}{\v {\hat h}}
\newcommand{\T}{\mathsf{T}}

\title{A model of sensory neural responses in the presence of unknown modulatory inputs \footnotemark}

\author{
\hbox to 5in{  
Neil C. Rabinowitz \hfil
Robbe L. T. Goris \hfil
Johannes Ball\'{e} \hfil
Eero P. Simoncelli}\\[2ex]
Howard Hughes Medical Institute,\\
Center for Neural Science\\
New York University\\
New York, NY  10003\\[1ex]
\texttt{eero.simoncelli@nyu.edu} \\
}

\nipsfinalcopy 

\begin{document}

\maketitle
\footnotetext{Submitted to NIPS 2014 (not accepted). This version has been slightly modified to correct some minor mistakes; the original submission is preserved as v1 in arxiv.}

\begin{abstract}
Neural responses are highly variable, and some portion of this variability arises from fluctuations in modulatory factors that alter their gain, such as adaptation, attention, arousal, expected or actual reward, emotion, and local metabolic resource availability.  Regardless of their origin, fluctuations in these signals can confound or bias the inferences that one derives from spiking responses.  Recent work demonstrates that for sensory neurons, these effects can be captured by a modulated Poisson model, whose rate is the product of a stimulus-driven response function and an unknown modulatory signal [1]. Here, we extend this model, by incorporating explicit modulatory elements that are known (specifically, spike-history dependence, as in previous models [2, 11]), and by constraining the remaining latent modulatory signals to be smooth in time.  We develop inference procedures for fitting the entire model, including hyperparameters, via evidence optimization [3], and apply these to simulated data, and to responses of ferret auditory midbrain and cortical neurons to complex sounds.  We show that integrating out the latent modulators yields better (or more readily-interpretable) receptive field estimates than a standard Poisson model. Conversely, integrating out the stimulus dependence yields estimates of the slowly-varying latent modulators.
\end{abstract}

\section{Introduction}

One of the great mysteries of neuroscience is how the brain manages to perform stable and useful computations using such apparently unreliable elements as neurons. One can present the same stimulus to a sensory neuron over and over again, and the number of spikes it produces will differ each time. While some of this variability arises from biophysical processes within neurons (e.g., synaptic transmission, diffusion processes within and across membranes), we have also known for a long time that neural responses are affected by a huge array of contextual state variables: arousal; attention; expected or actual reward; emotions; local metabolic resource availability; and the presence of stimulants, depressants or anesthetics. In turn, fluctuations in these variables can compound the measured variability of neural responses 
[e.g., 12, 13, 14].

Although these sources of variability are well known in the neuroscience community, many of them are difficult to control or measure. Experimentalists largely handle this variability by simple averaging.  In sensory systems, for example, stimuli are typically presented repeatedly for many trials; the results are then averaged, in order to estimate the stimulus-dependent component of the neural response. This approach has several major drawbacks. First, the need for many repeats is very resource-intensive. Second, it assumes that the variability can actually be removed by averaging, which need not be true. Third, there are no widely agreed-upon means for assessing the contribution of these unknown sources, before or after averaging. 

If we wish to study a non-stationary brain, then it behooves us to consider more explicitly how these unknown state variables manifest in neural responses. A more explicit model of their effect could, in principle, have two major benefits: it might allow us to directly estimate the contextual influences on neurons, and it could provide better methods to infer stimulus-response relationships. 

To first approximation, the effect of these state variables is {\em modulatory}: they alter the gain or responsivity of neurons, rather than directly cause the generation of spikes (Fig.\ 1). This idea has recently proven useful as a structural constraint for a model of the desired form. Goris et al [1] describe neural responses as arising from a Poisson process whose rate is a product of stimulus drive and a latent gain signal. This model provides an accurate  account of the variability of sensory neurons, attributing a substantial fraction to fluctuations in modulatory influences.  Here, we generalize this {\em modulated Poisson} (MoP) model, providing it with a more explicit stimulus-dependent component, and assuming that the unknown state variables driving the gain fluctuations change only slowly in time. We develop statistical inference procedures for extracting both the latent gain signal, as well as the parameters of the stimulus dependency.  We test these procedures on simulated data, and demonstrate the success of the model in explaining electrophysiological data obtained from auditory neurons in the anesthetized ferret.

We note that our formulation has similarities to previously published models. In form, this is an extension of the Generalized Linear Model (GLM, see below; [2]), but with the inclusion of a time-varying latent signal. Our model shares the general form of a latent state space model for neural data [15-19]. This class includes the Poisson Linear Dynamical System (PLDS) model [17]. Unlike the PLDS, which relies on autoregressive processes to characterise the latent signal, our formulation relies on Gaussian Processes. We also relax the constraint that the latent modulatory signal is passed through the same nonlinearity as the stimulus-response function, and rather consider it simply to be a positive-valued signal that has a multplicative effect.

An earlier version of these results was presented at the CoSyNe meeting in 2014.

\begin{figure}[h]
\begin{center}
\includegraphics[width=0.95\linewidth]{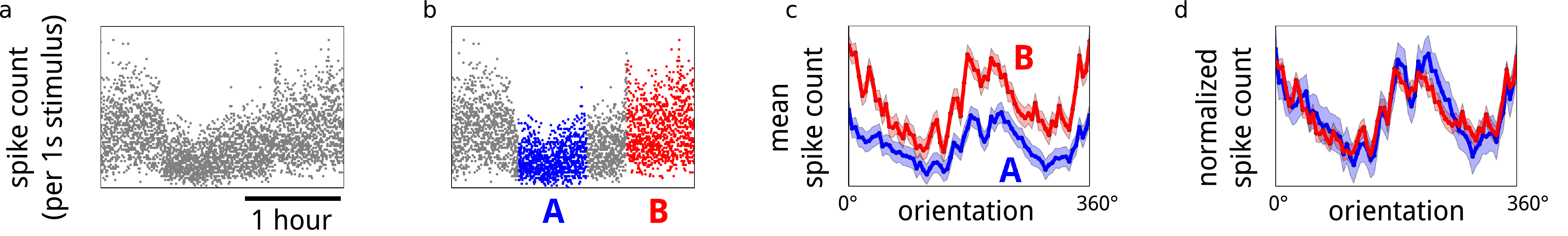} 
\end{center}
\caption{Neurons are subject to modulatory influences, which affect their spiking output. (a) Spike counts recorded from a real V1 neuron, in the anesthetized macaque, from [6]. Stimuli (oriented gratings) drive the firing of this neuron, but have dynamics on a timescale of seconds. Modulatory influences on this cell create slow non-stationarities that are visible in its spiking output. (b) Two epochs of low and high firing (labelled A and B).  (c-d) the stimulus selectivity of the neuron in these two epochs differs by a multiplicative gain factor.}
\end{figure}

\section{Modulated Poisson Model}

Our central abstraction is to build a generative model of neural responses that consists of three components. First, there are the controlled or measured inputs to a neuron. For a sensory neuron, this might be encapsulated as ``the stimulus''.  Second, there a slowly varying stochastic modulatory process, that acts as a multiplier on the instantaneous firing rates. This component is not directly observed. Third, there is the spike-generating point process (whose output is typically observed). 

For more detail, we begin with the classical Poisson framework for describing the stimulus-driven response of a sensory neuron (Fig.\ 2a). This asserts a simple generative model for producing spike counts within a set of time windows, and contains only the input-rate and point-process components. We assume here we have a collection of discrete time windows of identical duration, indexed by $t$, and that the spike rate within each of these windows, $\mu_t$, is a function, $F$, of the current stimulus, $\xt$ (or, perhaps, the recent history of stimulation, which can be encapsulated in $\xt$), and a set of input-response parameters, $\v k$:
\begin{equation}
    \mu_t = F( \xt; \v k )
\end{equation}
In turn, we assume that the spike count within each window, $y_t$, is drawn from a Poisson distribution with rate $\mu_t$. We use the vector notation $\v y$ to describe the set of scalars $y_t$. 

This class includes the Linear-Nonlinear-Poisson (LNP) models (where $\mu_t = F( \xt^\T \v k )$ ), the Generalized Linear Model (GLM), where the input $\xt$ encapsulates a range of known time-varying signals, such as the recent history of spike counts from the same and other neurons [2], and the Generalized Quadratic Model (GQM), where $F$ also includes quadratic operations on $\xt$ [7]. 

In the Modulated-Poisson (MoP) model [1], we assume that the firing rate is additionally modulated by a multiplicative interaction with a latent (unobserved) variable $g_t$:
\begin{equation}
    \mu_t = g_t \cdot F( \xt; \v k )
\end{equation}
We make two specific assumptions about the vector of time-varying modulator values, $\v g$. First, we assume it is positive, by writing $g_t = \exp(h_t)$. This choice is not unique: other nonlinear functions could also be used [10]. Second, we assume that it is slowly varying. The slow dynamic on $\v h = \left[ h_t \right]$ could take on any form appropriate to a given system (e.g.\ linear Gaussian, discrete Markov). Here, for tractability and generality, we assume that $\v h$ is multivariate Gaussian:
\begin{equation}
\label{prior.h}
\v h  \;\sim\;  \mathcal N \left( \v{0}, \Ch(\th) \right)
\end{equation}
where the smoothness is imparted by $\Ch(\th)$, the prior covariance matrix on $\v h$, via a set of hyperparameters $\th$. This structure is depicted in the schematic shown in Fig.\ 2b, and the graphical model shown in Fig.\ 2c. Below, we motivate a particular choice of form for $\Ch(\th)$.

\begin{figure}[h]
\begin{center}
\includegraphics[width=0.95\linewidth]{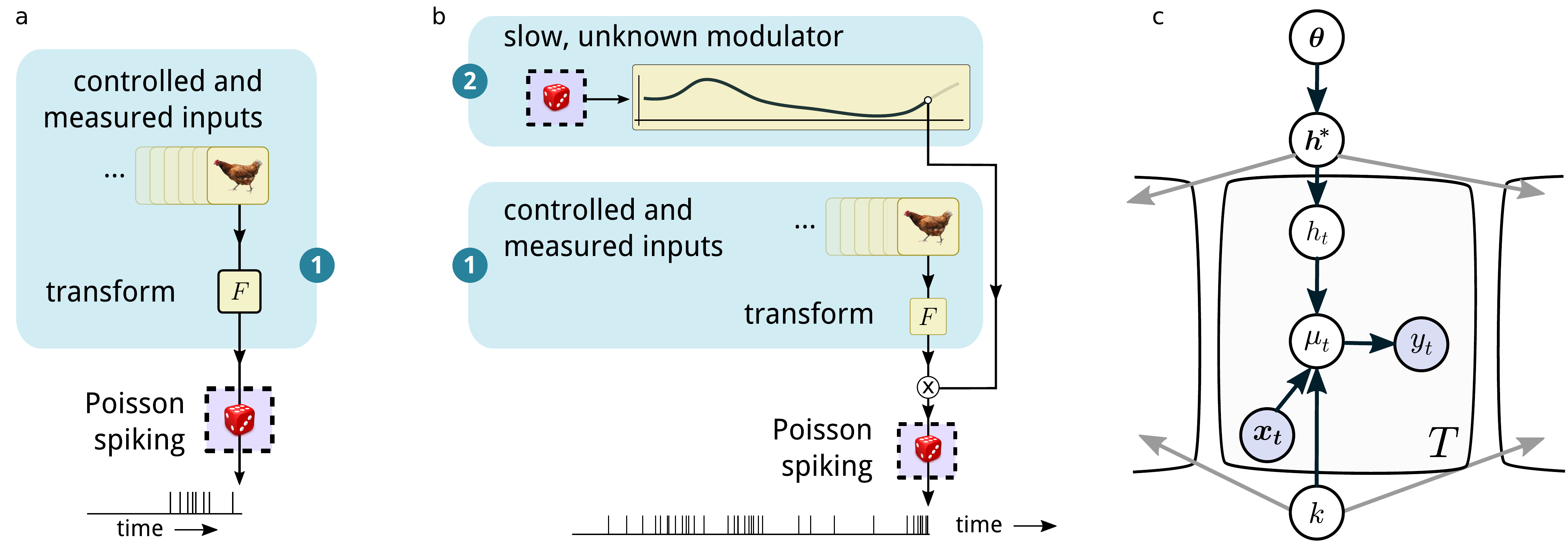} 
\end{center}
\caption{(a) The classical Poisson model, and (b) the Modulated-Poisson model for the generation of spikes by a neuron. The latter adds a structured stochastic gain modulation. (c) Graphical model representation of the MoP-model.}
\end{figure}

\section{Inference}

Suppose we have a set of neural data with stimuli $\v X = \left[ \xt \right]$ and observed spike counts $\v y$. Our goal is to infer the two unknown components of the model: the log-modulators $\v h$, and the parameters of the input-response relationship, $\v k$. In the case where the nonlinearity $F$ is the exponential function, these two stages can be merged. For generality, we maintain our central abstraction here, and split the inference into separate stages for each of these two components. 

More precisely, we approximate the joint posterior by $p(\v h, \th, \v k | \v X, \v y) \approx q_k(\v k | \v X, \v y) \cdot q_{h,\theta}(\v h, \th | \v X, \v y)$, and then solve iteratively for the two components. For brevity, we omit the subscripts on $q$ and the conditioning on $\v X$ and $\v y$ from the notation, and write these simply as $q(\v k)$ and $q(\v h, \th)$. Here, we shall remain agnostic to the form of the input-response relationship ($F$), and thus focus entirely on the modulator component. Further, rather than solving jointly for the modulator $\v h$ and its hyperparameters $\th$, we follow the approach of [3], and split this inference into two parts: first, we calculate a maximum-marginal-likelihood point estimate $\thhat = \operatorname*{arg\,max}_{\th} p( \v X, \v y | \th, \v k )$, and then we calculate a posterior on $\v h$ conditioned on $\thhat$. We concentrate on the latter problem first.

\subsection{Inference on $h$}
\label{section:inference.on.h}

Suppose we have an estimate of $q(\v k)$, and $\thhat$, a point estimate of $\th$. To solve for $\v h$ given these, we begin by writing the terms of the log joint, $\LJ = \log p(\v h, \v \theta, \v k; \v X, \v y)$ that depend on $\v h$, and then we condition on our current estimates of $q(\v k)$ and $\thhat$:
\begin{equation}
    \mathbb{E}_{q(\v k), \thhat}\left[ \LJ \right]  \;=\;  -\v \nuh^\T \exp(\v h) \+ \v y^\T \v h -\half \v h^\T \Cthhatinv \v h \+ \const
\label{LJ}
\end{equation}
where $\nuh$ is the vector of values $\hat \nu_t = \Ek{ F( \xt; \v k ) }$, i.e.\ the expected stimulus-driven component of the firing rate under our current estimate of $q(\v k)$, and the exponential is applied elementwise.

The number of elements of $\v h$ is equal to the number of time windows, $T$. This is typically very large. For example, an hour of continuous recording, with spike counts binned at 25~ms intervals has $T \approx 10^5$; at 5~ms resolution, this number exceeds $7 \times 10^5$. Performing inference on $\v h$ scales with $O(T^3)$, and is therefore intractable: the objective is expensive to compute, and the number of variables to optimize is far too large. We therefore reduce its dimensionality. Our strategy is to project $\v h$ into a low-dimensional space specified by the eigendecomposition of the prior covariance, $\v C$. Since we want $\v h$ to be slowly varying, the prior to be translation invariant, and the eigendecomposition and projection to be inexpensive, a natural choice is for $\v C$ to be circulant. In this way, the eigenvectors are sinusoids, the prior is diagonalized to a Fourier power spectrum (in this case, low-pass) which can be parameterized by $\th$, the transform can be quickly calculated through the FFT, and inference can proceed on the Fourier coefficients of $\h$ within the pass-band.

More specifically, we write: $\v C = \v Q^\T \v L \v Q$ as the eigendecomposition; $\v P$ as the $(T^* \times T)$ projection operator which excises the low-eigenvalue dimensions, i.e.\ those with low prior power; $\v P^\T$ as its pseudoinverse, which imputes missing coefficients as zero; and $\v R = \v P \v Q$ as the combined transform to the reduced space. We consider the DFT matrix $\v Q$ as an invertible real-valued transform, by concatenating the real and imaginary parts of the transformed vectors; since $\h$ is real-valued, the negative frequencies are redundant and can be omitted from the computation. We implement all matrix operations that involve $\v Q$ and $\v R$ via an orthonormalized, real-valued, real-signal FFT.

As a result, we perform inference on the length-$T^*$ vector $\hstar$, corresponding to a reduced Fourier-domain representation of the modulator, $\h$. The transform back to the time domain is given by $\h = \v R^\T \hstar = \v Q^\T \v P^\T \hstar$. This reduces the complexity of inference to $O(T \log T \,+\, (T^*)^3)$, corresponding to the cost of the FFT plus the cost of inference in reduced space.

In reduced Fourier space, the conditional log posterior on $\hstar$, its Jacobian and its Hessian become:
\begin{align}
    \LPhstar  
    \;&=\;  
        -\nuh^\T \exp\left( \v R^\T \hstar \right) 
        \+ \v{y}^\T \v R^\T \hstar  
        \-  \half  \hstar^\T \v \Lstarinv \hstar
    \label{LPh.objective}
    \\
    \jacob \LPhstar \hstar  
    \;&=\;  
        \v R \left( \v y - \v \mu \right)  
        \- \v \Lstarinv \hstar
    \label{LPh.jacobian}
    \\
    \hesscross \LPhstar \hstar {{\hstar}^\T}
    \;&=\;  
        -\v R \, \diag{\v \mu} \v R^\T  \- \v \Lstarinv
    \label{LPh.hessian}
\end{align}
where $\v {\mu} = \nuh \, \odot \, \exp( \v R^\T \hstar)$ is the firing rate, $\odot$ is the elementwise product, and the diagonal matrix $\Lstar = \v R \v C \v R^\T = \v P \v L \v P^\T$ is the projected prior covariance on $\hstar$. Since equation \eqref{LPh.objective} contains a sum of exponentials, a moment-matched exponential family form for $q(\hstar | \th)$ cannot be tractably normalized. However, since $\mu_t \ge 0$, the Hessian is convex wrt.\ $\hstar$, so we can solve for a Laplace approximation of the conditional posterior instead, $q(\hstar | \v \theta) = \mathcal{N} ( \hhatstar, \Lamstar )$. In principle, this reduced Fourier representation can be transformed back into the time domain, giving $ q(\v h | \th) = \mathcal{N} ( \v {\hat h}, \v \Lambda ) = \mathcal{N} (\v R^\T \hhatstar, \v R^\T \Lamstar \v R )$, but since $\v \Lambda$ is $(T \times T)$, it is typically too large to fit in memory. Sampling and conditional inference can nevertheless proceed with $q(\hstar)$ rather than $q(\v h)$.

In order for the FFT to be an efficient tool, the time windows should be regularly sampled. It is nevertheless straightforward to handle missing data at time points $t^\prime$, by simply setting $\hat {\nu}_{t^\prime} = y_{t^\prime} = 0$. In this way, imputed values of $h_{t^\prime}$ do not materially contribute to the likelihood terms in the objective and its derivatives, and are constrained only by the smoothness and shrinkage terms from the prior. This also provides a way of avoiding artefacts from circular boundary conditions imposed by the Fourier transform: we pad $\h$ with an additional $T$ values, for which the responses are unobserved. 

Finally, the first term of the Hessian is expensive to compute directly, even via the FFT implementation. However, we note that this matrix is a selection of rows and columns from $-\v Q \, \diag{\v \mu} \, \v Q^\T$, which, in turn, is a real-valued rearrangement of the real and imaginary parts of a complex-valued, Hermitian, circulant matrix, with entries constructed from the Fourier transform of $\v \mu$. This enables a quick construction.

\subsection{Inference on $\theta$}

We place a smoothness prior on $\h$. For a given neural recording, the degree of smoothness, i.e.\ the timescales over which the latent modulators vary, is itself unknown. We therefore build a hierarchical model, wherein the smoothness is controlled by a set of hyperparameters, $\th$. In the Fourier domain, this amounts to parameterizing the prior power spectrum -- i.e.\ the elements of the diagonal of $\v L$ -- as a function of $\th$. 

One potential form for a low-pass prior on $\v h$ is the zero-mean ALDf prior (Automatic Locality Determination in the frequency domain [3]). This parameterizes the elements of the diagonal of $\v L$ as a function of their respective Fourier frequencies, $f(i)$, and two hyperparameters $\th = \lbrace \sigma^2, \rho \rbrace$:
\begin{equation}
    \left[ \v L \right]_{ii}  \;=\;  \exp \left( -\frac {f(i)^2} {2 \sigma^2} \- \rho \right )
\end{equation}
In practice, there is considerable computational pressure to minimize the number of coefficients of $\hstar$ (see Section \ref{section.recovering}), which means we have to truncate this spectrum at some point. We find it more practical to construct a low-pass prior covariance from a more compact windowing function, such as a Blackman-Harris window, with hyperparameters $\th = \lbrace F_c, \rho \rbrace$:
\begin{equation}
    \left[ \v L \right]_{ii}  
        \;=\;  
        \left\{
            \begin{array}{lr} 
                e^{-\rho} \sum_{n=0}^{3} (-1)^n a_n \cos \left( \pi n (1 + f(i)/F_c) \right) & \;:\; f(i) \le F_c \\
                0 & \;:\; f(i) > F_c
            \end{array}
        \right.
\end{equation}
where $\v a = [0.35875, 0.48829, 0.14128, 0.01168]^\T$. Compared with the truncated ALDf prior, this has less flexibility in controlling the ripple in the estimate of $\h$. Nevertheless, the difference in fitted $\h$ values are typically very small for a modest speed improvement.

To choose values for the hyperparameters, we follow the approach of Park and Pillow [3, 4], and solve for the maximum marginal likelihood value of $\th$, a procedure also known as evidence optimization. The idea is to integrate out the values of $\h$, and maximize the marginal likelihood of the data as a function of $\th$. This gives a point estimate of $\th$, as $\thhat$. We refer the reader to these papers for a complete description of this process.

\subsection{Inference on $k$}

In parallel to the approach in Section \ref{section:inference.on.h}, we suppose we have an estimate of the modulators $q(\v h, \th)$, and wish to solve for the input-response parameters, $q(\v k)$. The terms of the log joint containing $\v k$ are:
\begin{align}
    \mathcal{LJ}  \;=\;  -F(\xt; \v k)^\T \exp(\h) \+ \v y^\T \log{F(\xt; \v k)} \+ \log p(\v k)  \+ \const
\end{align}
where the exponential and logarithm are applied element-wise. Next, we condition on our current estimate of $q(\v h, \th)$. To do this, we need only replace $\exp(\v h)$ with its expectation under $q(\v h, \th)$:
\begin{align}
\Ehth{ \exp(\v h) }
    \;=\;
        \exp \left( \hhat + \half \mathrm{diag} \, \v \Lambda \right)
    \;=\;
        \exp \left( \hhat + \half \left( (\v R^\T \Lamstar) \odot \v R^\T \right) \v 1 \right)
\end{align}
where the latter expression avoids explicitly constructing $\v \Lambda$. We calculate $\v R^\T \Lamstar$ via an IFFT on the columns of $\Lamstar$, and we build $\v R^\T$ by taking the IFFT of the columns of $\v P^\T$. Inference on $\v k$ then proceeds in a manner appropriate to the specific form of the function $F$, e.g.\ by maximizing $\Ehth{\LJ}$ wrt.\ $\v k$. 

In practice, we find that two to three iterations of solving for $q(\v h, \th)$ given $q(\v k)$, then $q(\v k)$ given $q(\v h, \th)$, are sufficient for convergence.

\section{Results}

\subsection{Simulations}

\subsubsection{Recovering the modulator}
\label{section.recovering}

\newcommand{\htrue}{\v {h_{\mathrm{true}}}}
\newcommand{\ktrue}{\v {k_{\mathrm{true}}}}
\newcommand{\thtrue}{\v {\theta_{\mathrm{true}}}}

We start by asking how well this inference procedure recovers underlying modulators when we know the ground truth. To answer this, we simulated a neuron's spiking response to a stimulus ensemble. We assumed a simple input: the model neuron's expected response to a stimulus of value $x$ was described by a smooth tuning function $f(x)$, which we discretized as a length-$K$ vector, $\v k$. We also subjected the model neuron to a slow gain fluctuation (Fig.\ 3a). We set the ground truth for $\ktrue$ and $\thtrue$, and sampled the modulator, $\htrue$, and spikes, $\v y$, from the generative model, for $T$ consecutive, pseudo-random presentations of the stimuli. We then performed complete inference of $\v k$, $\v h$, and $\v \theta$.

\begin{figure}[h]
\begin{center}
\includegraphics[width=0.95\linewidth]{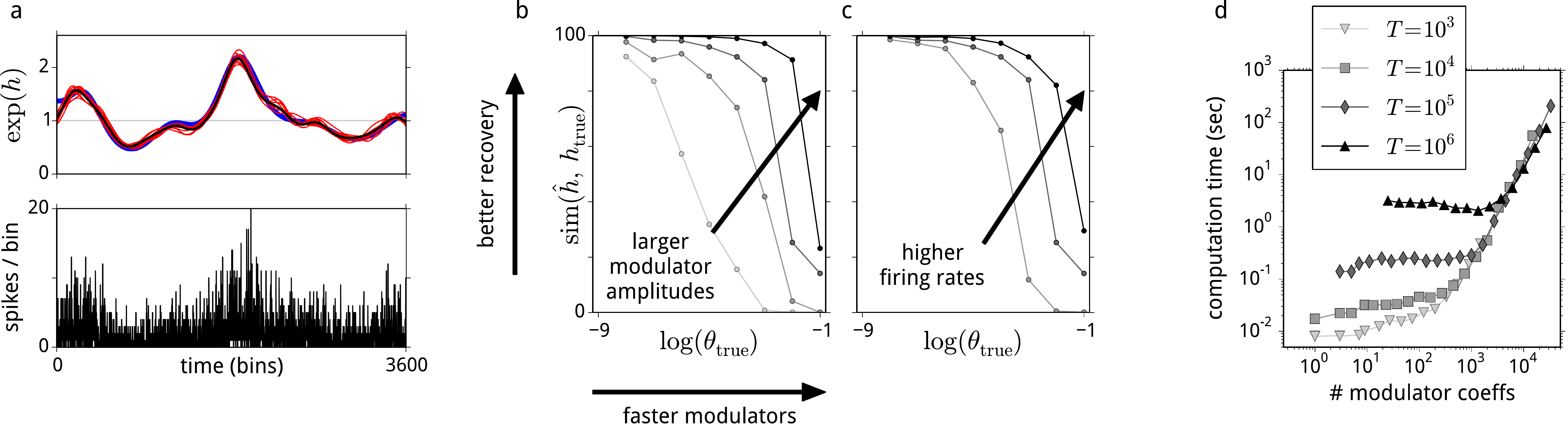} 
\end{center}
\caption{Simulated data, illustrating inference on the latent modulator. (a) Top: we sampled a modulator (blue) for a given $\th$, and applied it to a model neuron responding to a stimulus. From the spiking output (bottom), we inferred $q(\v h)$, which is overlaid on the top panel. Black: posterior mode, $\hhat$; red: samples from the posterior, $q(\v h)$. (b)-(c): Quality of recovery of $\htrue$, measured as $\mathrm{sim}(\hhat, \htrue) = 100 \times (1 - \mathrm{Var}(\hhat - \htrue)/\mathrm{Var}(\htrue))$, improves with several factors: increasing the cutoff frequency of the imposed modulators (abscissae); increasing the standard deviation of the imposed modulator ((b); factors of $1/4$, $1/2$, $1$, and $4$); and increasing the firing rate ((c); factors of $1/4$, $1$, and $4$. (d) The computational cost of computing $q(\v h | \th)$ as a function of $\th$ (abscissa, measured via the length of $\hstar$), and the number of time windows (symbols). }
\end{figure}

We reasoned that the quality of the modulator estimate would depend on how much data was available to estimate each phase of the modulator, as well as the quality of that data. We tested this in three ways. First, for slow modulations, the posterior mode of the recovered modulator, $\hhat$ was very close to the ground truth, $\htrue$; but when the imposed modulators were too fast, and approached the sampling resolution of the time windows themselves, the fit quality declined. Second, when we reduced the magnitude of the fluctuations in $\htrue$, the estimates $\hhat$ captured the slower modulations only (and $\thhat$ was more conservative than $\thtrue$). Finally, we observed the same result when we decreased the average firing rate (Fig.\ 3b-c). These results reflect an important feature of the inference procedure: by maximizing the marginal likelihood of $\th$, we learn the fastest timescale of modulation that the data are able to provide evidence for.

The inference on $\v h$ scales as $O(T \log T + (T^*)^3)$, and is thus constrained by two factors (Fig.\ 2d). First, as the number of observations grows, the FFT operations become more costly, regardless of the value of $\th$. Second, as the number of modulator coefficients (determined by $\th$) grows, the optimization of $\hstar$ becomes more expensive. This cost appears to dominate as the hyperparameter admits higher frequencies. Since computation time is always limited, it is necessary to set some upper bound on the maximum length of $\hstar$, and hence on $\th$, for each dataset. Our experiments suggest that keeping the number of modulator coefficients below about 1000--2000 is a reasonable goal. This, however, means that any modulations at higher frequencies cannot be tractably learned using this algorithm.

\subsubsection{Recovering the input-response relationship}
\label{section:unbiasing}

Even if we are not interested in recovering the modulatory signal per se, including it in the model can yield a substantial improvement in estimates of the input-response parameters, $\v k$.

To demonstrate this, we sampled from a simple generative model for the spiking responses of auditory cortical neurons, based on the Generalized Linear Model (GLM, [2]; Fig.\ 3). The auditory stimulus was a Dynamic Random Chord (DRC) ensemble as described in [5]. Briefly, this is a collection of simultaneously-presented tones; the levels of each tone were drawn independently every 25 ms from a fixed distribution. We simulated neurons with a fixed set of parameters: a linear spectro-temporal receptive field (STRF) that was localized in time and frequency, a pointwise exponential nonlinearity, and a spike feedback term that modulated the firing rate for a period of up to 200 ms after every spike. In addition, we added gain modulations of different kinds. We then fitted GLMs to the simulated spike trains, assuming an ALDs prior on the STRF, and an ALDsf prior on the spike history term [3]. We also added latent modulators to the models. For brevity, we refer to the GLMs fitted without the latent modulators as P-GLMs, as the only stochastic source in these models is the conditionally Poisson point process, while we refer to GLMs with the latent modulators as MoP-GLMs, as they additionally have a stochastic modulator process.

In the absence of slow modulations, i.e.\ under conditions of stationary gain, the P-GLMs recovered the ground truth STRFs and spike history terms with a high degree of accuracy (Fig.\ 4a). However, when there was either a slowly varying gain (Fig.\ 4b) or a  slow monotonic decrease in gain over time (Fig.\ 4c), the fitted spike history term for the P-GLMs included a near-constant positive offset. This artefact reflects the only means the model has to capture a slowly changing gain state: to smooth the last few seconds' spike counts as a proxy measure of the modulator signal, and leverage this to change the predicted firing rate in the next time bin. Indeed, when we increased the speed of the imposed gain modulations, the estimated spike feedback kernels differed wildly from the ground truth (Fig.\ 4d). In all cases, switching to a MoP-GLM removed these biases, and recovered more accurate estimates of the spike-feedback kernels.

\begin{figure}[h]
\begin{center}
\includegraphics[width=0.95\linewidth]{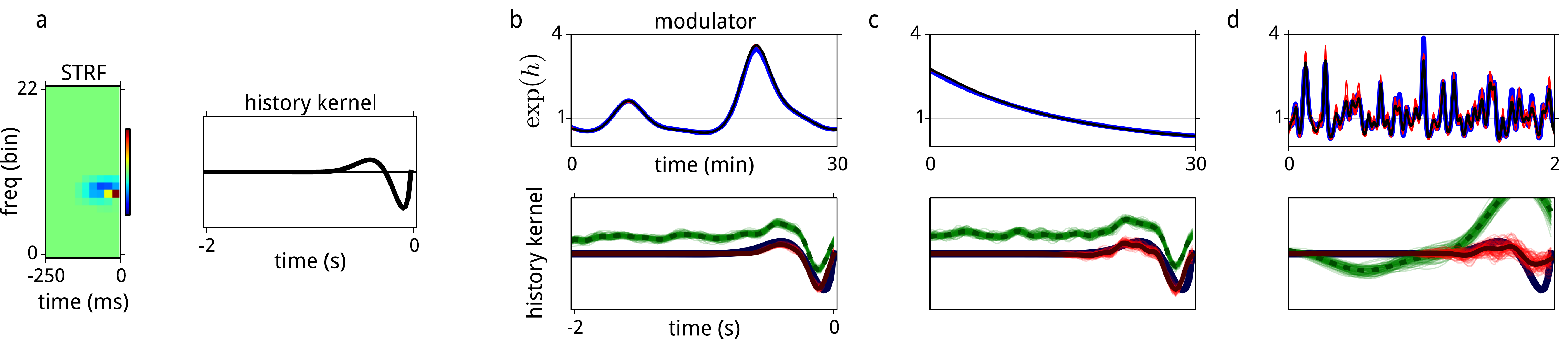} 
\end{center}
\caption{Simulated data, illustrating inference on $\v k$ with and without the latent modulator. (a) Ground truth receptive field and spike-feedback kernel used to generate spikes according to a GLM forward model. (b)-(d) Top row: modulators applied to simulated neurons (blue); estimates $\hhat$ from the MoP-GLM model showing excellent recovery (black/red). Bottom: comparison of P-GLM (green) and MoP-GLM estimates (black/red) of spike-history kernels. Without accounting for the modulator, the inference on $\v k$ is biased.}
\end{figure}

\subsection{Real neural data}

\subsubsection{Improved predictive power}
\label{section:improved}

To test the modulated-Poisson model in practice, we fit GLMs to real spiking data obtained from extracellular recordings of auditory neurons in the primary auditory cortex and midbrain of anesthetized ferrets. The details of the data collection are described in [5, 9]. We fitted both P-GLMs and MoP-GLMs, as per the previous section, to approximately an hour of spiking responses to random chord stimulation from each of 339 recorded neurons. These data were recorded under typical non-stationary physiological conditions; we note, for instance, that the depth of anesthesia in such preparations is known to vary over time (though was not quantified here).

We trained the GLMs on 80\% of the data. We constructed the held-out 20\% test set by selecting random 250 ms contiguous snippets from each recording, with a minimum of 250 ms training data between each test snippet. By setting $\nu_{t^\prime} = y_{t^\prime} = 0$ for these snippets, inference on the training data simultaneously yielded predictive distributions on the modulators for the test data. Since the test snippets were very short in duration compared with the fastest tractable latent modulator timescale, the values of $\v h$ were almost constant over each test snippet. In particular, we emphasize that while the MoP-GLMs have more fitted parameters than the P-GLMs, these extra parameters were fit exclusively to the training dataset: we introduced no new latent variables to describe the held-out data. 

Since ground truth is unknown for these datasets, we simply compared the likelihoods on the test datasets for the P-GLM and MoP-GLM model fits. For all 339 neurons, the MoP-GLMs produced better predictions of the held-out data than the P-GLMs. The improvement scaled with the amount of non-stationarity estimated in the spike train (Fig.\ 5a). Moreover, the model-estimated value of the modulator hyperparameters, $\thhat$, appeared optimal for prediction: when we fixed the modulator time constants to values higher or lower than the learned values, predictions were measurably worse (Fig.\ 5b). 

\begin{figure}[h]
\begin{center}
\includegraphics[width=0.95\linewidth]{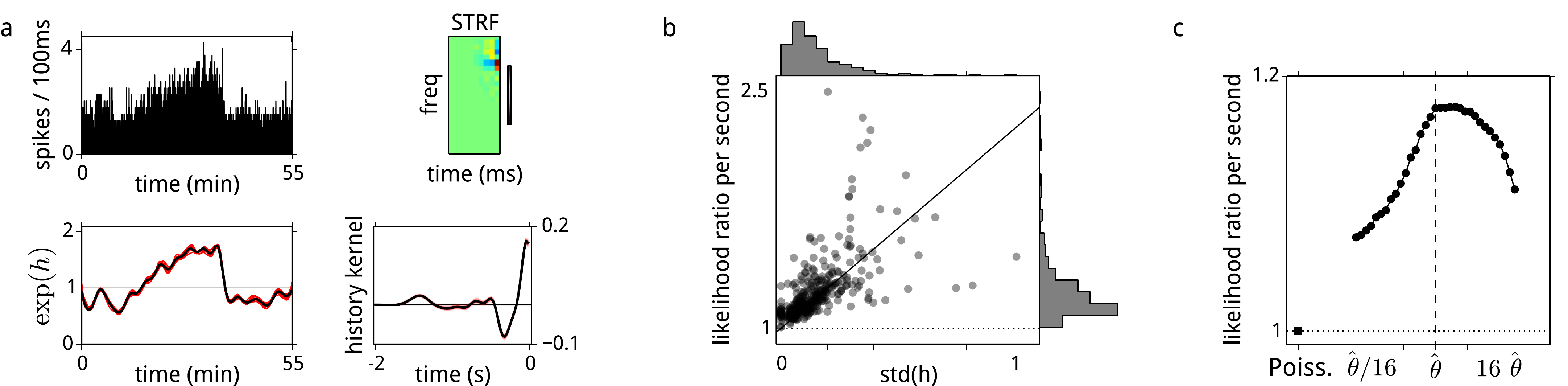} 
\end{center}
\caption{Real data, showing the improved predictive power of a model which includes a latent modulator. (a) Example multi-unit neural responses from auditory cortex [5]. Top left: spike counts. Remaining panels show fits of the components of MoP-GLM to this cell. (b) Improved predictions on held-out data from 339 units, measured as the likelihood ratio per second between the MoP-GLM and the P-GLM. The difference was greatest for cells with more estimated modulation ($r = 0.6$). (c) Median improvement of MoP-GLM over Poisson-GLM across the population, when the cutoff frequency was set to a value higher or lower than the learned optimum. This shows the local sensitivity to mis-estimation of $\thhat$. }
\end{figure}

\section{Discussion}

We developed a modulated-Poisson model, that includes a smooth, latent modulatory signal, which is combined multiplicatively with a stimulus-driven firing rate, as well as an inference procedure for estimating all parameters given noisy neural data. Our application of this model to simulated data indicates that the inference recovers an underlying modulator of the neuron's firing rate, so long as this modulator varies on a sufficiently slower time scale than the windows over which spikes are counted. On the other hand, if we exclude the modulator from our models, our simulations show that this can lead to substantial biases in the estimation of stimulus-response parameters. When we apply this model to real data from the auditory midbrain and cortex of the ferret, we find that cross-validated predictions of neural responses improve. We believe that the tools we have developed here offer a principled solution to a long-standing problem in experimental neuroscience: how to build analysis for a non-stationary brain.

\subsubsection*{Acknowledgments}

We thank Tony Movshon, Andrew King, and Jan Schnupp for sharing electrophyiological data. 

\subsubsection*{References}

\small{
[1] Goris, Robbe L.T., Movshon, J.A., \& Simoncelli, E.P. (2014). Partitioning neuronal variability. {\it Nature Neuroscience} {\bf 17}(6):858--65.

[2] Pillow, J. W., Shlens, J., Paninski, L., Sher, A., Litke, A. M., Chichilnisky, E. J., \& Simoncelli, E. P. (2008). Spatio-temporal correlations and visual signalling in a complete neuronal population. {\it Nature}, {\bf 454}(7207):995--999.

[3] Park, M., \& Pillow, J.W. (2011). Receptive field inference with localized priors. {\it PLoS Comput Biol} {\bf 7}(10):e1002219.

[4] Park, M., \& Pillow, J.W. (2013). Bayesian inference for low rank spatiotemporal neural receptive fields. {\it Advances in Neural Information Processing Systems}, pp. 2688--2696. Cambridge, MA: MIT Press.

[5] Rabinowitz, N.C., Willmore, B.D., Schnupp, J.W., \& King, A.J. (2012). Spectrotemporal contrast kernels for neurons in primary auditory cortex. {\it The Journal of Neuroscience}, {\bf 32}(33):11271--11284.

[6] Graf, A.B., Kohn, A., Jazayeri, M., \& Movshon, J.A. (2011). Decoding the activity of neuronal populations in macaque primary visual cortex. {\it Nature Neuroscience}, {\bf 14}(2):239--245.

[7] Park, I.M., Archer, E.W., Priebe, N., \& Pillow, J.W. (2013). Spectral methods for neural characterization using generalized quadratic models. {\it Advances in Neural Information Processing Systems}, pp. 2454--2462. Cambridge, MA: MIT Press.

[8] Sahani, M., \& Linden, J.F. (2003). Evidence optimization techniques for estimating stimulus-response functions.{\it Advances in neural information processing systems}, pp. 317--324.

[9] Rabinowitz, N.C., Willmore, B.D., King, A.J., \& Schnupp, J.W. (2013). Constructing noise-invariant representations of sound in the auditory pathway. {\it PLoS Biology}, {\bf 11}(11), e1001710.

[10] Paninski, L. (2004). Maximum likelihood estimation of cascade point-process neural encoding models. {\it Network: Computation in Neural Systems}, {\bf 15}(4), 243--262.

[11] Truccolo, W., Eden, U. T., Fellows, M. R., Donoghue, J. P. \& Brown, E. N. (2004). A point process framework for relating neural spiking activity to spiking history, neural ensemble and extrinsic covariate effects. {\it J. Neurophysiol.} {\bf 93}: 1074--1089.

[12] Tomko, G.J. \& Crapper, D.R. (1974). Neuronal variability: non-stationary responses to identical visual stimuli. 
{\it Brain Res.} {\bf 79}:405-–418.

[13] Tolhurst, D.J., Movshon, J.A. \& Thompson, I.D. (1981). 
The dependence of response amplitude and variance of cat visual cortical neurones on stimulus contrast. 
{\it Exp. Brain Res.} {\bf 41}:414-–419.

[14]  Churchland, M.M. et al. (2010). 
Stimulus onset quenches neural variability: a widespread cortical phenomenon. 
{\it Nat. Neurosci.} {\bf 13}:369-–378.

[15]  Kulkarni, J. E. \& Paninski, L. (2007).
Common-input models for multiple neural spike-train data.
{\it Network: Computation in Neural Systems}, {\bf 18}: 375--407.

[16]  Paninski, L. et al. (2010).
A new look at state-space models for neural data. 
{\it Journal of Computational Neuroscience}, {\bf 29}: 107--126.

[17]  Macke, J. H. et al. (2011).
Empirical models of spiking in neural populations. 
{\it Advances in Neural Information Processing Systems}, pp. 1350--1358.

[18]  Vidne, M. et al. (2012).
Modeling the impact of common noise inputs on the network activity of retinal ganglion cells. 
{\it Journal of Computational Neuroscience}, {\bf 33}: 97--121. 

[19]  Archer, E. W., Koster, U., Pillow, J. W. \& Macke, J. H. (2014).
Low-dimensional models of neural population activity in sensory cortical circuits. 
{\it Advances in Neural Information Processing Systems}, pp. 343--351.

}

\end{document}